\documentstyle[eqsecnum,twocolumn,aps]{revtex}

\begin{document}
\draft
\preprint{}
\title{Properties of the massive Thirring model from the
$\mathbf{XYZ}$ spin chain}
\author{Marko Kolanovi\'{c}, Silvio Pallua, and Predrag Prester}
\address{Department of Theoretical Physics, PMF, University of Zagreb\\
Bijeni\v{c}ka c. 32, 10001 Zagreb, Croatia}
\maketitle
\begin{abstract}
We consider here the massive Thirring model regularized with the $XYZ$ spin
chain. We numerically calculate the mass ratios of particles which lie
in the discrete part of the spectrum and obtain results in
accordance with the DHN formula and in disagreement with recent 
calculations in the
literature based on the numerical Bethe ansatz and infinite momentum
frame methods. We also analyze the short distance behavior of these
states and evaluate the conformal dimensions. This paper, taken together
with the previous one for the sine-Gordon model, confirms the duality
relation between two models formulated by Klassen and Melzer [Int.
J. Mod. Phys. A {\bf 8}, 4131 (1993)].
\end{abstract}

\narrowtext

\section{Introduction}
\label{sec:intro}

The massive Thirring model (MTM) and sine-Gordon mo\-del (SGM) are
important as a testing laboratory for understanding ideas proposed for
other more complicated field theories.

In this paper we propose to calculate certain physical quantities for
the MTM by performing an explicit diagonalization of its lattice
regularization with the $XYZ$ spin chain. As the
first task we want to calculate the masses of breathers. The
previous calculations have been based on the semiclassical method
\cite{DHN}, on factorized scattering theory \cite{Zam1}, or on
the Bethe ansatz method \cite{BeTh,Kor,Lut,Lusch,WeiSch,JaNeWi,Vega}.

The additional interest in avoiding the previously mentioned
assumptions is due to recent criticism \cite{FuOg,FuSeYa,FuHi}
[the authors claim that there is only one breather in the whole
attractive region, and with different mass than the
Dashen-Hasslacher-Neveu (DHN) formula predicts]. The same authors 
challange also the well-known duality relation between the MTM and 
SGM \cite{Col,Man,DeGW,GoSt,KlMelz}. The precise meaning and extent
of this equivalence was formulated by Klassen and Melzer
\cite{KlMelz} (notice that the models are not equivalent when they 
have a finite size in space).

One important criticism relates to the use of the so-called
string conjecture. Indeed, violations of this conjecture are
observed in the literature \cite{string}. Despite the fact that, at
least untill now, it was not known that these violations affect any
relevant results, it would be desirable to have a calculation which
does not rely on the string conjecture.

It is for this reason that we want to treat the MTM without using
the above-mentioned assumptions. Our approach will be based on
direct numerical diagonalization of the $XYZ$ spin chain which is a
lattice regularization of the MTM \cite{Lut,Lusch}. This method is
suited for analyses of low discrete states in the spectrum, but
becomes less and less effective when we go to higher states. Such
an approach was used in the literature for other problems, e.g.,
conformal unitary models perturbed by some relevant operator
\cite{HenSal,Hen,Hen2}.

We also intend to calculate conformal dimensions of operators creating
breather states. There are conjectured values for them
\cite{KlMelz}. By explicit calculation we confirm this conjecture
for the first breather but get different results for the second breather.

Recently \cite{PP99} we have performed a similar calculation for the
SGM. The regularization in this case was the $XXZ$ spin chain in a
transverse field. The results on the masses of breathers and conformal
dimensions agree as statements on relation of two models would
suggest, so it maybe considered also as an independent check of
the SGM-MTM correspondence \cite{KlMelz}.

\section{MTM as a massive perturbation of the Gaussian model}
\label{sec:gauss}

The MTM is a $(1+1)$-dimensional field theory of a Dirac spinor field $\psi$,
defined classically by the Lagrangian
\begin{equation} \label{lmt}
{\cal L}_{MTM}=i\bar{\psi}\gamma^{\mu}{\partial_{\mu}}\psi-\lambda\bar{\psi}
\psi-\frac{g}{2}(\bar{\psi}\gamma_{\mu}\psi)(\bar{\psi}\gamma^{\mu}\psi).
\end{equation}
Here $\lambda$ is a dimensionful parameter which sets the mass scale
in a theory which is conformaly invariant when $\lambda=0$. However,
although $\lambda$ enters Eq. (\ref{lmt}) as a (bare) mass, its mass
dimension $d_{\lambda}$ is not equal to 1 but is determined from
the (nontrivial) anomalous dimension of the field $\bar{\psi}\psi$.
The dimensionless coupling constant $g$ is scale invariant (vanishing
beta function), but it is not uniquely defined due to the existence of
different regularizations of the (conserved) current
$\bar{\psi}\gamma^{\mu}\psi$. Correspondingly, there is at least a
one-parameter family of definitions of $g$. Our definition will be
the same as the one used by Coleman \cite{Col} (Schwinger
definition). We shall find it more convenient to use the parameter 
$\beta$ related to $g$ with
\begin{equation} \label{dua}
\frac{4\pi}{\beta^2}=1+\frac{g}{\pi}\, .
\end{equation}
Here $\beta$ is the dimensionless coupling constant from the dually
related SGM.

In \cite{KlMelz} it was shown that the MTM can be viewed as a perturbed
conformal field theory (CFT) when the second term in Eq. (\ref{lmt}) 
is treated as a (massive) perturbation. We will now repeat here some 
results of their analyses relevant for our discussion.

An unperturbed theory $\lambda=0$ (approached in the UV limit) is the
Thirring model which is a CFT with central charge $c=1$ and an
operator algebra generated by
\begin{equation} \label{lfg}
L_{f}=\left\{ V_{m,n}|m\in 2{\bf Z},n\in {\bf Z}
 \mbox{ or }m\in 2{\bf Z}+1,n\in {\bf Z}+1/2 \right\}
\end{equation}
where $V_{m,n}(x)$ are primary fields with conformal dimensions
\begin{equation} \label{cod}
(\Delta_{m,n},\bar{\Delta}_{m,n})=\left(\frac{2\pi}{\beta^2}\left(
\frac{m\beta^2}{8\pi}+n\right)^{\! 2},\frac{2\pi}{\beta^2}\left(
\frac{m\beta^2}{8\pi}-n\right)^{\! 2}\right) ,
\end{equation}
where we used duality relation (\ref{dua}). From Eq. (\ref{cod}) we
can read off the scaling dimensions and (Lorentz) spin of $V_{m,n}$:
\begin{eqnarray} \label{scd}
d_{m,n}&=&\Delta_{m,n}+\bar{\Delta}_{m,n}=\frac{m^2 \beta^2}{16\pi}
+\frac{4\pi n^2}{\beta^2}\, , \\ s_{m,n}&=&\Delta_{m,n}-\bar{\Delta}_{m,n}
=mn \, . \nonumber
\end{eqnarray}
A whole operator algebra is generated by a quartet of fields $V_{\pm
1,\pm 1/2}$, which are connected to the fundamental spinor fields
$\psi$, $\bar{\psi}$ by
\begin{equation} \label{pvc}
\psi\longleftrightarrow \left(\begin{array}{c} V_{1,1/2} \\
V_{-1,1/2} \end{array}\right), \qquad
\bar{\psi}\longleftrightarrow \left(\begin{array}{c} V_{1,-1/2} \\
V_{-1,-1/2} \end{array}\right) .
\end{equation}
Now one supposes that Hilbert space of the full (perturbed) theory
is isomorphic to that of the unperturbed one. From operator product
algebra (OPA) it follows that the (properly normalized) perturbing
operator in the MTM (\ref{lmt}) is
\begin{equation} \label{pop}
\bar{\psi}\psi=V^{(+)}_{2,0}\equiv\frac{1}{2}(V_{2,0}+V_{-2,0})\,,
\end{equation}
which means that $\lambda$ has mass dimension $d_{\lambda}=2-d_{2,0}
=2-\beta^2/4\pi$. From the condition of relevancy of the
perturbation, i.e., $d_{\lambda}>0$, we obtain Coleman's bound
$\beta^2<8\pi$ ($g>-\pi/2$). Also, from Eqs. (\ref{lfg}) and
(\ref{pop}) we can see that the MTM has a $\tilde{U}(1)\times Z_2\times
\tilde{Z}_2$ internal symmetry group. The $\tilde{U}(1)$ acts as
$V_{m,n}\rightarrow e^{i\alpha n}V_{m,n}$, while $Z_{2}$ and
$\tilde{Z}_2$ are generated by $R:V_{m,n}\rightarrow V_{-m,n}$ and
$\tilde{R}:V_{m,n}\rightarrow V_{m,-n}$, respectively.

\section{Spin chain regularization of the MTM}
\label{sec:chain}

It was argued a while ago \cite{Lut,Lusch} that the MTM on a
cylinder with proper (antiperiodic) boundary conditions (B.C.'s)
possesses spin chain regularization given by the $XYZ$ spin chain 
defined by the Hamiltonian
\begin{equation} \label{xyz}
H_{XYZ}=H_{XXZ}-h\sum_{n=1}^{N}\left(\sigma_{n}^{x}\sigma_{n+1}^{x}
-\sigma_{n}^{y}\sigma_{n+1}^{y}\right) ,
\end{equation}
where $\sigma^{x,y,z}$ are Pauli matrices, $N$ is an even integer, and
$H_{XXZ}$ is the Hamiltonian of the $XXZ$ spin chain:
\begin{equation} \label{xxz}
H_{XXZ}=\sum_{n=1}^{N}\left(\sigma_{n}^{x}\sigma_{n+1}^{x}+\sigma_{n}^{y}
\sigma_{n+1}^{y}+\Delta\sigma_{n}^{z}\sigma_{n+1}^{z}\right), 
\end{equation}
where $-1<\Delta<1$ [we also use standard parametrization $\Delta=-
\cos\gamma$, so $\gamma\in(0,\pi)$]. In Eqs. (\ref{xyz}) and 
(\ref{xxz}) sector-dependent B.C.'s should be used:
\begin{equation} \label{boc}
\sigma_{N+1}^{x,y}=\sigma_{1}^{x,y}(-1)^{N/2}C \, ,\qquad
\sigma_{N+1}^{z}=\sigma_{1}^{z}\, ,
\end{equation}
where
\begin{equation} \label{cdf}
C=\prod_{n=1}^{N}\sigma^{z}_{n}\, .
\end{equation}
From results of Ref. it \cite{Alc1} follows that the $XXZ$ chain 
(\ref{xxz}) with B.C. (\ref{boc}) gives in the continuum limit a CFT
with a space of states equal to that of $L_{f}$ in Eq. (\ref{lfg}), 
where
\begin{equation} \label{bgr}
\beta=\sqrt{8(\pi-\gamma)}\, .
\end{equation}
That leads us to the conjecture that, aside from irrelevant
corrections,
\begin{equation} \label{per}
\sigma_{n}^{x}\sigma_{n+1}^{x}-\sigma_{n}^{y}\sigma_{n+1}^{y}
\propto V_{2,0}^{(+)}
\end{equation}
in the continuum limit.

Now, the continuum limit is obtained letting $N\to\infty$ and $h\to 0$,
but at the same time keeping fixed the scaling parameter $\tilde{\mu}$:
\begin{equation} \label{tmh}
\tilde{\mu}\equiv hN^{d_{\lambda}}=hN^{2-\beta^{2}/4\pi}=
hN^{2\gamma/\pi} .
\end{equation}
In this limit, the mass gaps of the $XYZ$ chain are expected to 
satisfy a scaling law
\begin{equation} \label{mgs}
\tilde{m}_{i}=h^{1/d_{\lambda}}\tilde{G}_{i}(\gamma,\tilde{\mu})
=h^{\pi/2\gamma}\tilde{G}_{i}(\gamma,\tilde{\mu})\, .
\end{equation}
The scaling parameter $\tilde{\mu}$ is connected to $L$ (space extension
of continuum theory, i.e., MTM). For our purposes it is enough to
know that $\tilde{\mu}\to\infty$ ($\tilde{\mu}\to 0$) corresponds to
$L\to\infty$ ($L\to 0$), respectively.

\section{Mass spectrum}
\label{sec:mass}

Our goal here is to calculate the mass ratios of particles in the
MTM in the $L\to\infty$ limit using the connection with the $XYZ$ 
spin chain (\ref{xyz}). First we must numerically calculate the mass
gaps of the spin chain for finite $N$ and $h$. Then we must make a 
continuum limit, i.e., take $N\to\infty$ and $h\to 0$, keeping
$\tilde{\mu}$ fixed. Finally we should make a $L\to\infty$, i.e.,
$\tilde{\mu}\to\infty$, limit. In practice, it is preferable to do
the following \cite{HenSal,Hen,Hen2}: first take $N\to\infty$ with
$h$ fixed and afterwards extrapolate $h\to 0$. A difference is that
in the latter case one does $\tilde{\mu}\to\infty$ before $h\to 0$.
These limits are performed using the BST extrapolation method 
\cite{BulSt,HenSchutz}.

We numerically diagonalized Hamiltonian (\ref{xyz}) for up to 16
sites using the Lanczos algorithm. We are interested in the masses,
so we only need the zero-momentum sector. We should note here that in
Ref. \cite{Lusch} it was shown that true space translations are
generated not by an ordinary translation operator on the spin chain,
but by its square. From this and the fact that Hamiltonian (3.1)
commutes with the operator $C$, Eq. (3.4), it follows that we can 
break the Hamiltonian in the momentum-zero sector into four sectors named 
$0^\pm$, $\pi^\pm$, where $0,\pi$ is macroscopic momentum and $\pm$ 
denotes eigenvalue of $(-1)^{N/2}C$ (which can only
be $\pm1$ because $C^2=1$). We considered a number of values of 
coupling constant in the attractive regime ($g>0$, i.e., $\Delta>0$).
The structure of the spectrum is in agreement with the DHN
prediction, i.e., we obtain: vacuum, first breather ($B1$), second 
breather ($B2$) (when it exists), and
``continuum'' in $0^+$; fermion ($F$) and ``continuum'' in $0^-$; 
antifermion ($\bar{F}$) and ``continuum'' in $\pi^-$; ``continuum'' 
starting with $FF$ and $\bar{F}\bar{F}$ in $\pi^+$. Names for the
particle states and $FF$, $\bar{F}\bar{F}$ continuum will be 
confirmed by results for the mass ratios. But even we couldn't make an 
extrapolation (because of the poor scaling in the $\tilde{\mu}\to
\infty$ limit) for the lowest ``continuum'' state in $0^+$ for values 
of $g$ where the DHN formula predicts that it should be of $B1\,B1$ 
type, its scaling law in the $\tilde{\mu}\to0$ limit clearly shows 
that its scaling dimension is the one we expect for the $B1\,B1$ 
lowest continuum state, i.e., $d_{4,0}$. We should mention also 
that spectra in $0^-$ and $\pi^-$ are exactly degenerated which 
means that the $F$ and $\bar{F}$ mass gaps are equal
even on the lattice, which was not the case in similar analyses of
the SGM in \cite{PP99}.

In Figs \ref{mg3}--\ref{mg75} we present numerical results for the
scaled gaps $\tilde{G}_i$ for four states: fermion ($F$), first
breather ($B1$), second breather ($B2$), and lowest state in
the $FF$ continuum ($C$). This is of course a check of the scaling
relation (\ref{mgs}). Finally, partially extrapolated mass ratios
\begin{equation} \label{rtmr}
\tilde{r}_{a}(\Delta,h)=\lim_{\stackrel{N\to\infty}{h\;\mathrm{fixed}}}
\frac{\tilde{m}_{a}}{\tilde{m}_{F}}=\lim_{\stackrel{N\to\infty}{h\;
\mathrm{fixed}}}\frac{\tilde{G}_{a}}{\tilde{G}_{F}}\, ,\qquad a\in\{
B1,B2,C\}
\end{equation}
and fully extrapolated mass ratios of the first breather
\begin{equation} \label{fer}
\tilde{r}_{B1}(\Delta)=\lim_{h\to 0}\tilde{r}_{a}(\Delta,h),
\end{equation}
are given in Tables. \ref{r3}--\ref{r9}, together with the DHN
predictions \cite{DHN} and predictions of Fujita {\em et al}.
\cite{FuOg,FuSeYa}. Final extrapolation $h\to 0$ was possible only
for the first breather because scaling of the second breather is
worse and asks for a larger $N$ (probably $N\ge 24$). One can see 
that our results strongly confirm DHN and reject Fujita {\em et al}.

\section{UV (conformal) limit of particle states}
\label{sec:conf}

Let us now turn our attention to the opposite, i.e., UV, limit of our
results for the $XYZ$ spin chain. We mentioned in Sec. \ref{sec:chain}
that it obtained when $\tilde{\mu}\to 0$. From conformal
perturbation theory we expect the scaling relation
\begin{equation} \label{tmch}
\tilde{m}_{a}=h^{\pi/2\gamma}\left[ 2\pi\zeta d_{a}
\tilde{\mu}^{-\pi/2\gamma}+\tilde{H}_{a}(\gamma,\tilde{\mu})\right] ,
\end{equation}
where $d_a$ is the scaling dimension of the state $a$, and $\zeta$
is a well-known normalization factor,
\begin{displaymath}
\zeta=\frac{2\pi\sin\gamma}{\gamma}\, .
\end{displaymath}
From Eq. (\ref{tmch}) we can obtain the scaling dimensions of the particle
states $F$, $B1$, and $B2$ from the condition that $\tilde{H}_a$
should be less singular than $\tilde{G}_a$. Our results are given in
Table \ref{our}. They differ from those conjectured in \cite{KlMelz}
only for the second breather, which has scaling dimension equal to that
of the first breather. These results are in agreement with those in
\cite{PP99} for the SGM. This result for the scaling dimension of the
second breather was analyticaly confirmed in \cite{FRT99} using an
extension of nonlinear integral equation method (NLIE). In Figs. 
\ref{mh3}--\ref{mh75} we show the numeric results for reduced scaling
functions, where we used values from Table \ref{our} for the scaling
dimensions.

\section{Conclusion}

We have calculated in this paper the masses of breather states and
the anomalous dimensions of related operators for the MTM using spin
chain regularization. This is a direct numerical calculation
independent of assumptions such as the semiclassical approximation
\cite{DHN}, factorized scattering theory \cite{Zam1}, or Bethe ansatz
method \cite{BeTh,Kor,Lusch}. On the other hand, in a series of
papers based on numerical calculation within the Bethe ansatz method
\cite{FuSeYa} or using the infinite momentum frame technique 
\cite{FuOg}, different results have been claimed. Our
calculation confirms the conventional spectrum. In addition we
calculate the anomalous dimensions of operators creating breather
states. It agrees with conjecture in \cite{KlMelz} for the first
breather but disagrees for the second breather. This result is
consistent with the previous calculation for the sine-Gordon model,
i.e., consistent with equivalence relation between the two models
\cite{KlMelz}.
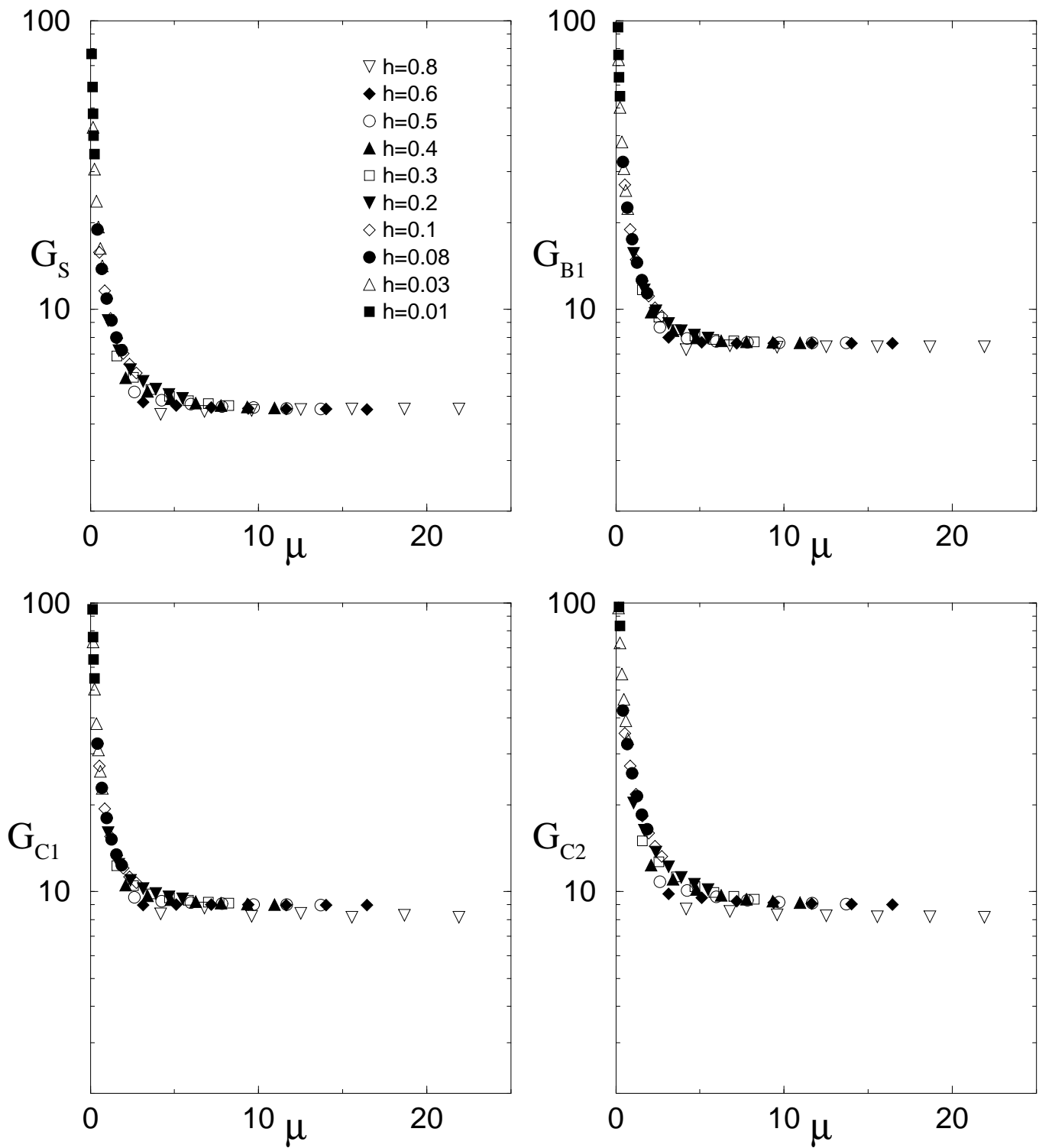
\begin{figure}
\caption{Scaling functions $\tilde{G}_{a}(\beta,\mu)$ for the
isolated gaps ($S$, $B1$) plus two lowest ``continuum'' gaps ($C1$,
$C2$) of the Hamiltonian (\ref{xyz}) at $\Delta=0.3$ (or
$\beta^2=10.13$, $g=0.76$). For this value of the coupling constant
the DHN formula predicts the existence of one breather. The legend
in the upper left figure applies to all figures in this article.}
\label{mg3}
\end{figure}
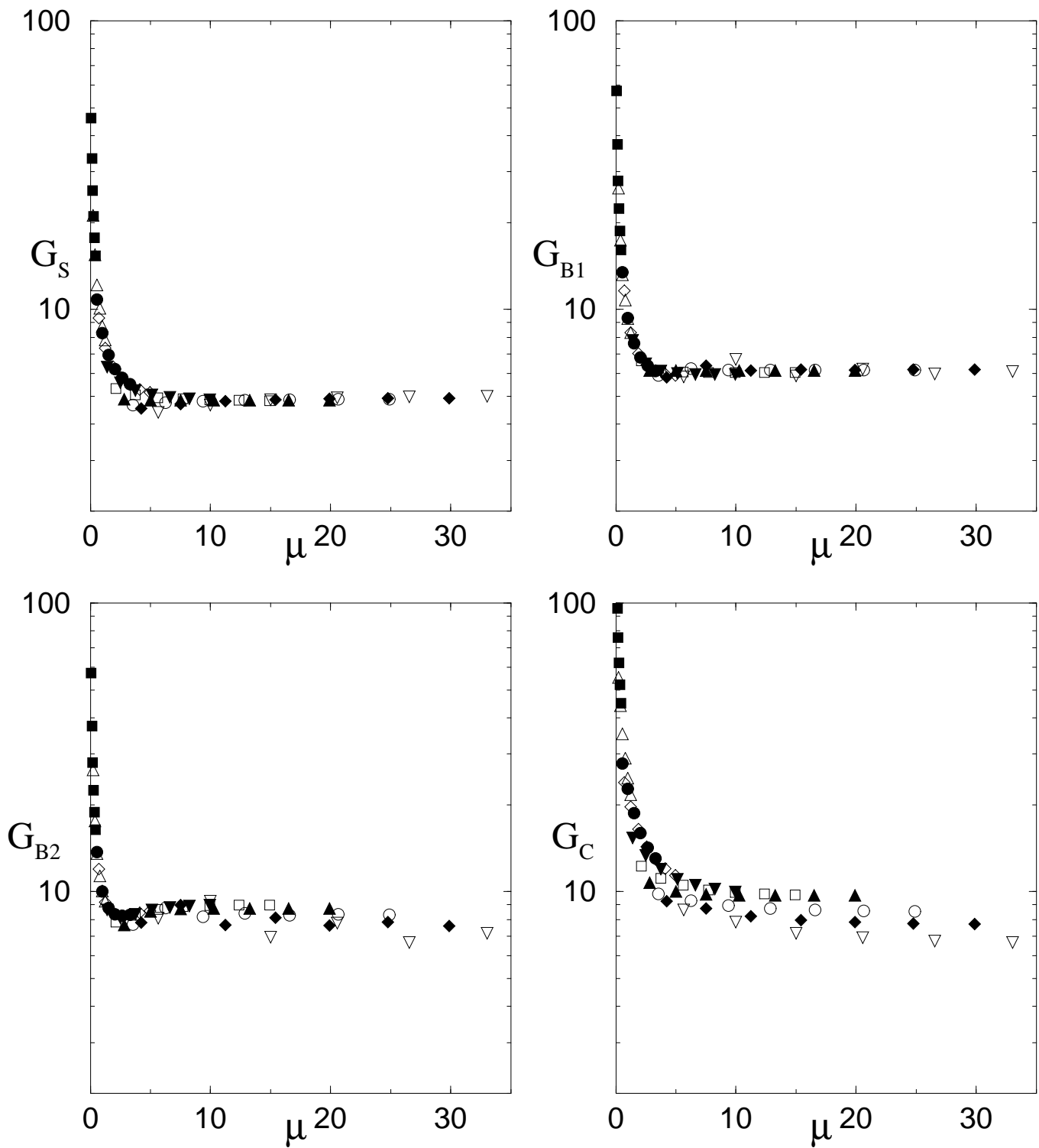
\begin{figure}
\caption{Scaling functions $\tilde{G}_{a}(\beta,\mu)$ for the
isolated gaps ($S$, $B1$, $B2$) plus lowest ``continuum'' gap ($C$) 
of the Hamiltonian (\ref{xyz}) at $\Delta=0.6$ (or $\beta^2=7.42$, 
$g=2.18$). The DHN formula predicts now the existence of two
breathers.}
\label{mg6}
\end{figure}
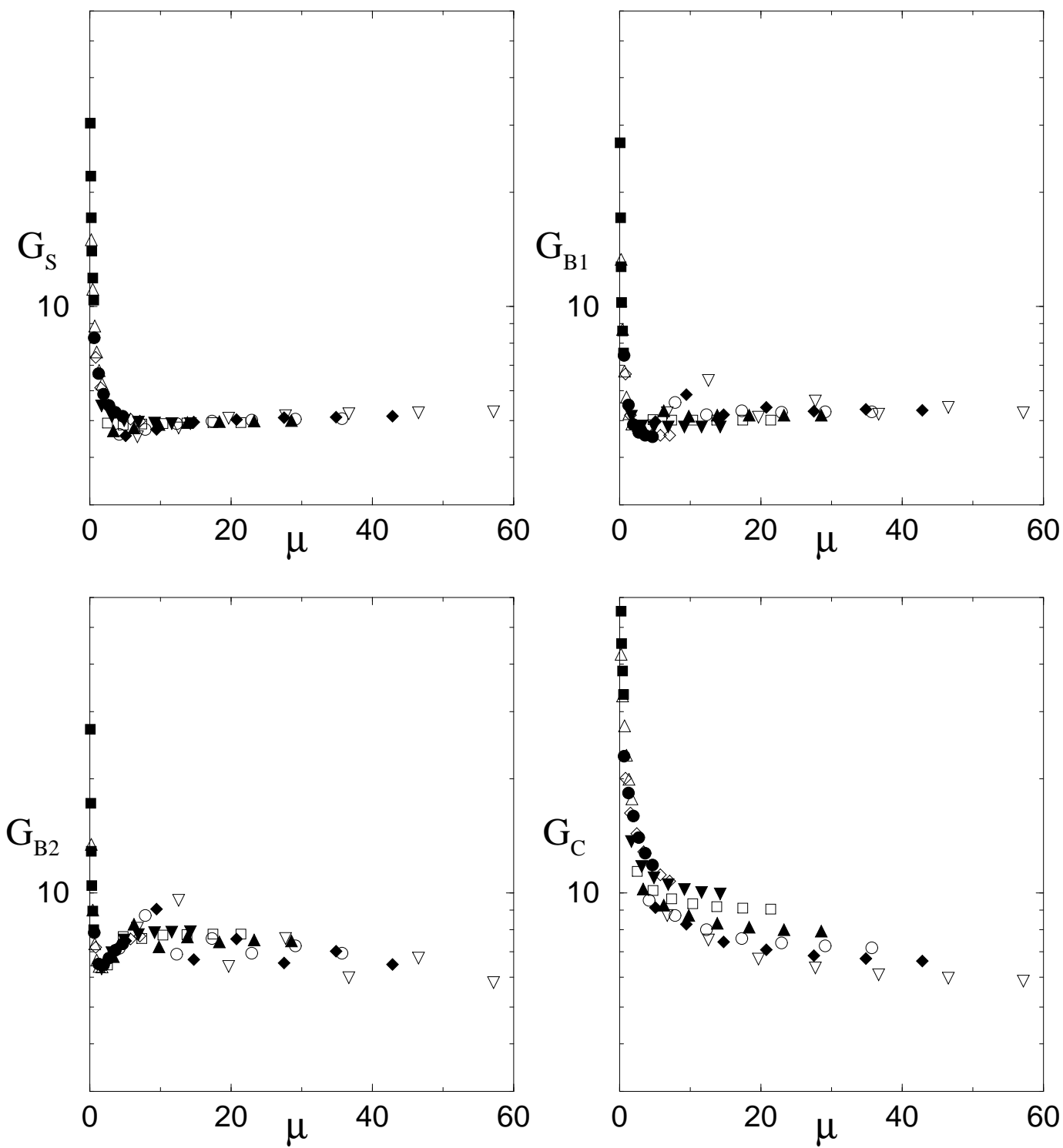
\begin{figure}
\caption{The same as Fig. \ref{mg6} but now for $\Delta=0.75$ (or
$\beta^2=5.78$, $g=3.69$).}
\label{mg75}
\end{figure}
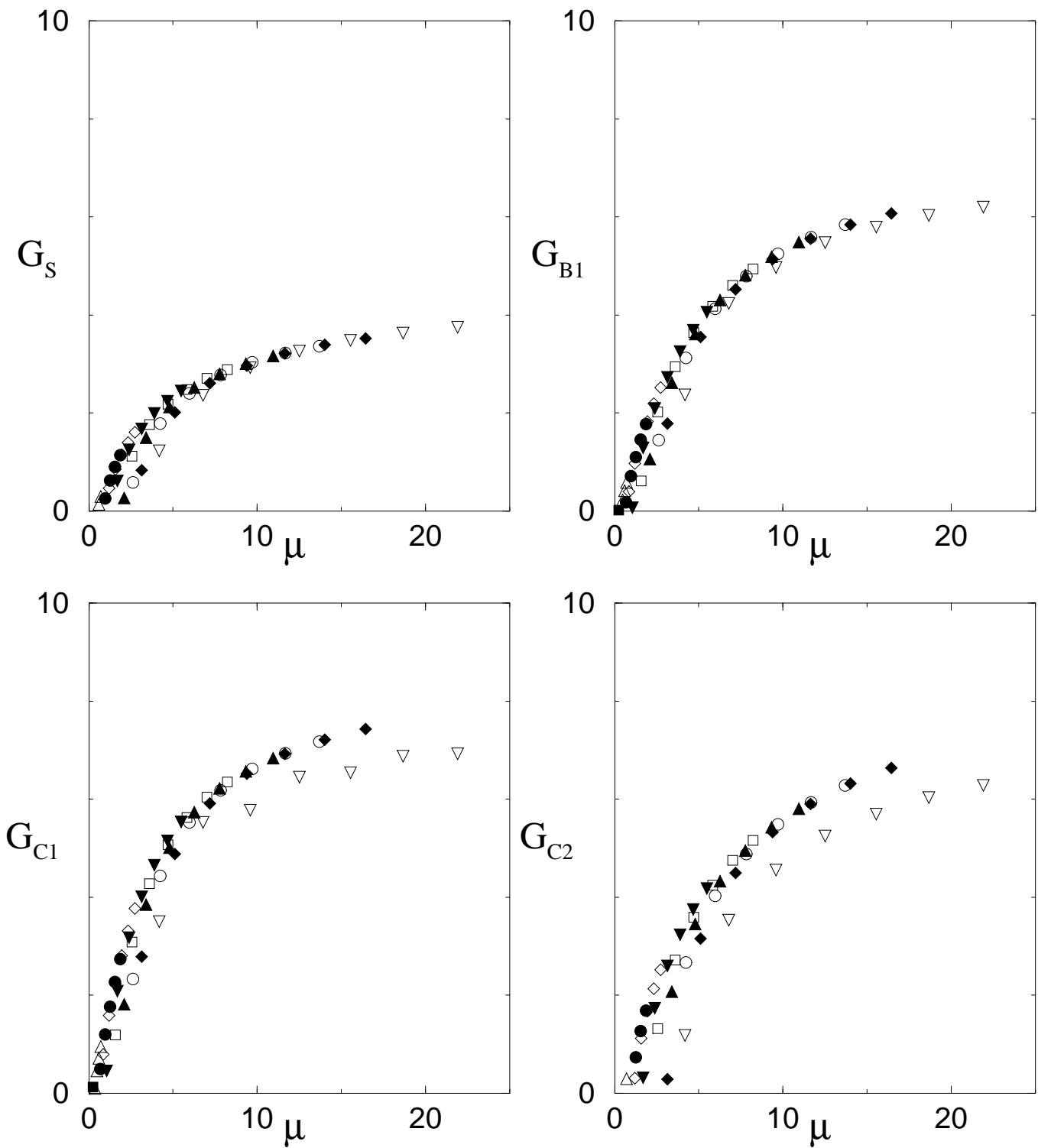
\begin{figure}
\caption{Reduced scaling functions $\tilde{H}_{a}(\beta,\mu)$ at
$\Delta=0.3$ (or $\beta^{2}=10.13$). The legend is the same as in
Fig. \ref{mg3}.}
\label{mh3}
\end{figure}
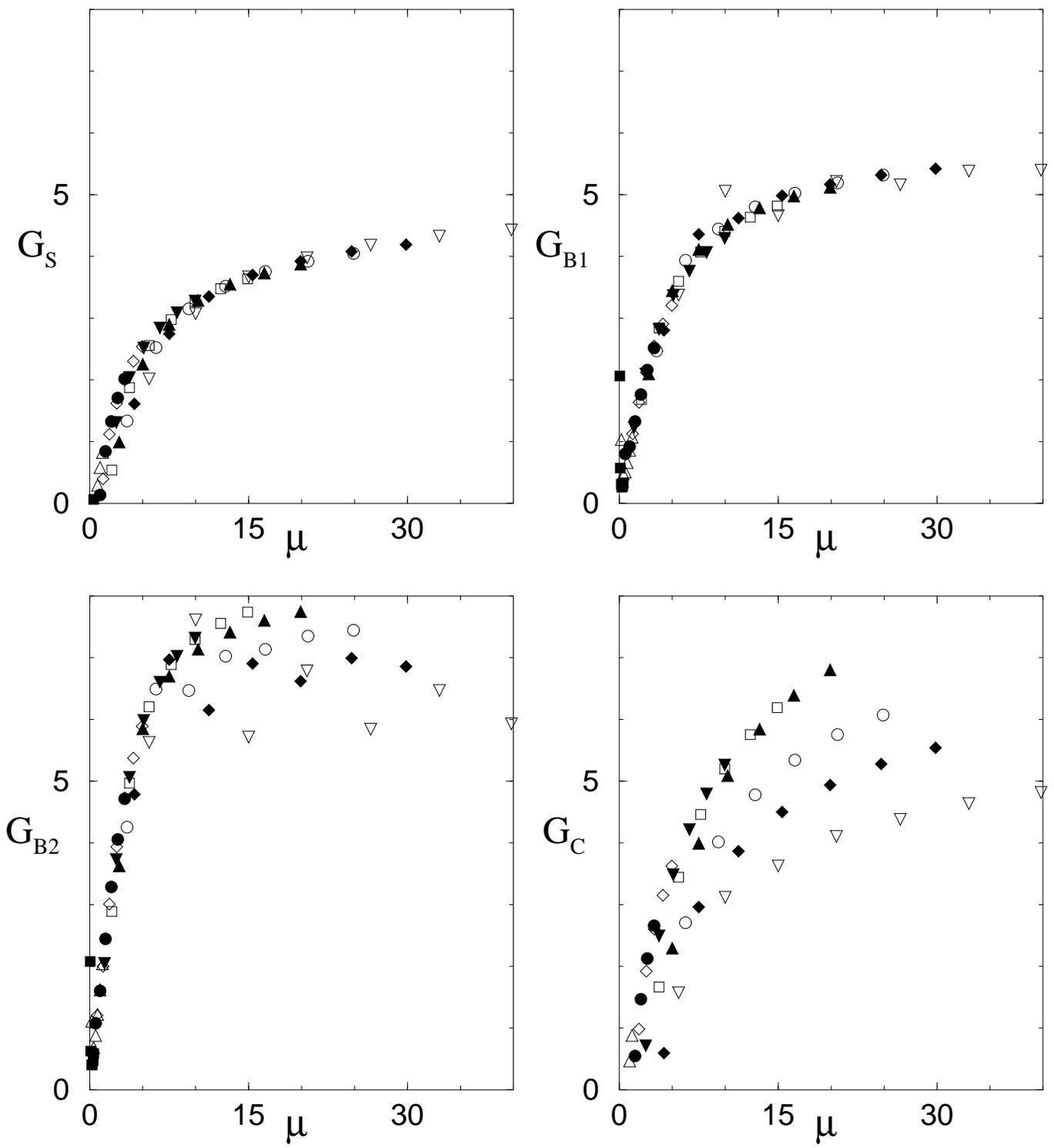
\begin{figure}
\caption{The same as Fig. \ref{mh3} but now for $\Delta=0.6$ (or
$\beta^2=7.42$).}
\label{mh6}
\end{figure}
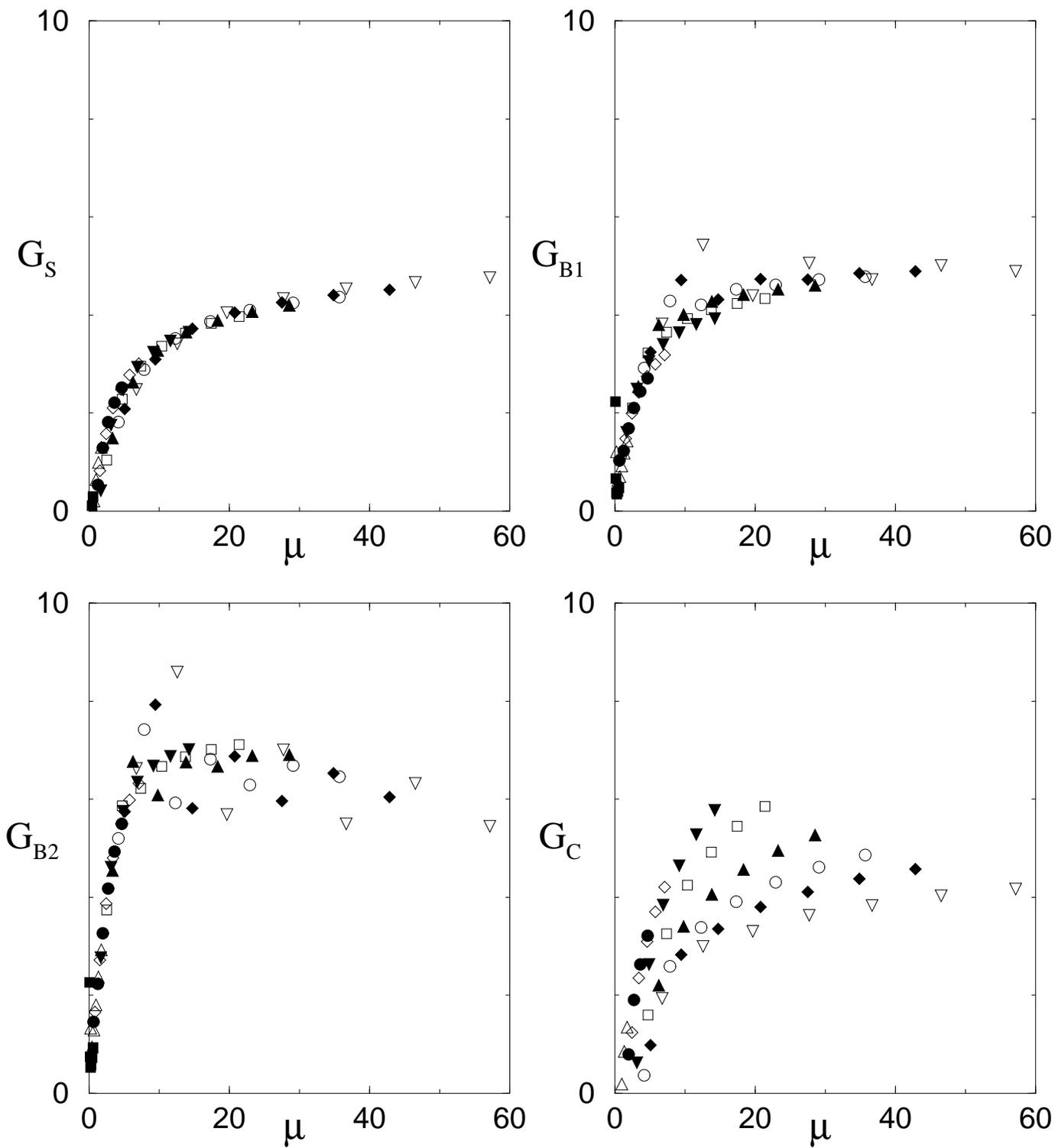
\begin{figure}
\caption{The same as Fig. \ref{mh3} but now for $\Delta=0.75$ (or
$\beta^2=5.78$).}
\label{mh75}
\end{figure}
\clearpage
\widetext
\begin{table}
\caption{Estimates for the mass gap ratios $\tilde{r}_{a}$ as a
function of $h$ at $\Delta=0.3$ ($\beta^{2}=10.13$, $g=0.76$). In
this regularization soliton and antisoliton gaps are exactly
degenerated. We also added the DHN prediction (only one breather
for this value of the coupling constant) and the prediction of
Fujita {\em et al.} (only one breather for all $g>0$). The numbers
in parentheses give the estimated uncertainty in the last given digit.}
\begin{tabular}{cllllllllll}
 & \multicolumn{7}{c}{$h$} & & \\ \cline{2-8}
\raisebox{1.5ex}[0pt]{$\tilde{r}_{a}$} & \multicolumn{1}{c}{0.8}
& \multicolumn{1}{c}{0.6} & \multicolumn{1}{c}{0.5} &
\multicolumn{1}{c}{0.4} &
\multicolumn{1}{c}{0.3} & \multicolumn{1}{c}{0.2} &
\multicolumn{1}{c}{0.1} &
\raisebox{1.5ex}[0pt]{$h\to 0$} & \raisebox{1.5ex}[0pt]{DHN} &
\raisebox{1.5ex}[0pt]{Fujita} \\ \hline
$B1$ & 1.6341\,(3) & 1.7007\,(4) & 1.718\,(1) & 1.730\,(3) &
1.734\,(8) & 1.74\,(2) & 1.67\,(6) & 1.747\,(6) & 1.745 & 1.777 \\
$C1$ & 1.786\,(7) & 2.0013\,(5) & 2.000\,(2) & 2.001\,(3) & 1.98\,(1)
& 2.00\,(2) & 2.07\,(5) &  & 2.000 & 2.000 \\
$C2$ & 1.797\,(2) & 2.0011\,(8) & 1.999\,(2) & 2.001\,(6) & 2.00\,(1)
& 2.00\,(3) & 1.93\,(8) &  & 2.000 & 2.000 \\
\end{tabular}
\label{r3}
\end{table}
\begin{table}
\caption{The same as Table \ref{r3} but now for $\Delta=0.6$
($\beta^{2}=7.42$, $g=2.18$).}
\begin{tabular}{cllllllllll}
 & \multicolumn{7}{c}{$h$} & & \\ \cline{2-8}
\raisebox{1.5ex}[0pt]{$\tilde{r}_{a}$} & \multicolumn{1}{c}{0.8}
& \multicolumn{1}{c}{0.6} & \multicolumn{1}{c}{0.5} &
\multicolumn{1}{c}{0.4} &
\multicolumn{1}{c}{0.3} & \multicolumn{1}{c}{0.2} &
\multicolumn{1}{c}{0.1} &
\raisebox{1.5ex}[0pt]{$h\to 0$} & \raisebox{1.5ex}[0pt]{DHN} &
\raisebox{1.5ex}[0pt]{Fujita} \\ \hline
$B1$ & 1.187\,(6) & 1.2443\,(6) & 1.2587\,(3) & 1.26491105\,(3) &
1.2638\,(5) & 1.254\,(2) & 1.240\,(9) & 1.24\,(2) & 1.223 & 1.337 \\
$B2$ & 1.2\,(1) & \multicolumn{1}{c}{-} & 1.694\,(2) & 1.807020\,(8)
& 1.8753\,(8) & 1.913\,(4) & 1.89\,(2) &  &
1.935 & 2.000 \\
$C$ & 1.29\,(1) & 1.536\,(4) & 1.734\,(2) & 1.99998\,(2) & 2.003\,(2)
& 2.00\,(1) & 1.99\,(4) &  & 2.000 & 2.000 \\
\end{tabular}
\label{r6}
\end{table}
\begin{table}
\caption{The same as Table \ref{r3} but now for $\Delta=0.75$
($\beta^{2}=5.78$, $g=3.69$).}
\begin{tabular}{cllllllllll}
 & \multicolumn{7}{c}{$h$} & & \\ \cline{2-8}
\raisebox{1.5ex}[0pt]{$\tilde{r}_{a}$} & \multicolumn{1}{c}{0.8}
& \multicolumn{1}{c}{0.6} & \multicolumn{1}{c}{0.5} &
\multicolumn{1}{c}{0.4} &
\multicolumn{1}{c}{0.3} & \multicolumn{1}{c}{0.2} &
\multicolumn{1}{c}{0.1} &
\raisebox{1.5ex}[0pt]{$h\to 0$} & \raisebox{1.5ex}[0pt]{DHN} &
\raisebox{1.5ex}[0pt]{Fujita} \\ \hline
$B1$ & \multicolumn{1}{c}{-} & 1.027\,(4) & 1.030\,(1) & 1.0255\,(4)
& 1.0112\,(2) & 0.9870\,(4) & 0.948\,(2) & 0.91\,(2) & 0.905 & 1.052
\\ 
$B2$ & 1.0\,(4) & 1.21\,(8) & 1.35\,(2) & 1.485\,(6) & 1.5716\,(3) &
1.6219\,(9) & 1.641\,(7) &  & 1.614 & 2.000 \\
$C$ & 1.528\,(3) & 1.25\,(1) & 1.360\,(4) & 1.553\,(4) & 1.803\,(2) &
2.005\,(4) & 1.97\,(2) &  & 2.000 & 2.000 \\
\end{tabular}
\label{r75}
\end{table}
\begin{table}
\caption{The same as Table \ref{r3} but now for $\Delta=0.9$
($\beta^{2}=3.61$, $g=7.79$).}
\begin{tabular}{cllllllllllll}
 & \multicolumn{9}{c}{$h$} & & \\ \cline{2-10}
\raisebox{1.5ex}[0pt]{$\tilde{r}_{a}$} & 
\multicolumn{1}{c}{0.6} & \multicolumn{1}{c}{0.5} &
\multicolumn{1}{c}{0.4} &
\multicolumn{1}{c}{0.35} &\multicolumn{1}{c}{0.3} &
\multicolumn{1}{c}{0.25} &
\multicolumn{1}{c}{0.2} & \multicolumn{1}{c}{0.15} &
\multicolumn{1}{c}{0.1} &
\raisebox{1.5ex}[0pt]{$h\to 0$} & \raisebox{1.5ex}[0pt]{DHN} &
\raisebox{1.5ex}[0pt]{Fujita} \\ \hline
$B1$ & 0.82\,(1) & 0.821\,(8) & 0.795\,(2) &
0.779\,(2) & 0.758\,(1) & 0.7356\,(5) & 0.7083\,(4) & 0.6763\,(4) &
0.63250\,(7) & 0.52\,(4) & 0.521 & 0.668 \\
$B2$ & 0.9\,(2) & 1.0\,(2) & 1.13\,(6) & 1.18\,(4)
& 1.18\,(1) & 1.202\,(8) & 1.1964\,(9) & 1.187\,(2) & 1.163\,(2)
&  & 1.005 & 1.336 \\
$C$ & 0.99\,(1) & 1.010\,(8) & 1.166\,(8) & 1.218\,(8) &
1.285\,(8) & 1.366\,(6) & 1.487\,(8) & 1.70\,(1) & 1.987\,(8) &
& 2.000 & 2.000 \\
\end{tabular}
\label{r9}
\end{table}
\narrowtext
\begin{table}
\caption{Scaling dimensions of particle states in the MTM as
conjectured from our numerical results.}
\begin{tabular}{lcc}
State & Operator & Scaling dimension \\
\hline \rule[-2mm]{0mm}{6mm}
Fermion & $V_{1,1/2}$ & $\frac{\beta^{2}}{16\pi}+
\frac{\pi}{\beta^{2}}$ \\ \rule[-2mm]{0mm}{6mm}
Antifermion & $V_{1,-1/2}$ & $\frac{\beta^{2}}{16\pi}+
\frac{\pi}{\beta^{2}}$ \\ \rule[-2mm]{0mm}{6mm}
First breather & $V_{2,0}^{(-)}$ & $\frac{\beta^{2}}{4\pi}$ \\ 
\rule[-2mm]{0mm}{6mm}
Second breather & $V_{2,0}^{(+)}$ & $\frac{\beta^{2}}{4\pi}$ \\
\rule[-2mm]{0mm}{6mm}
$\psi_1\psi_2$ continuum & $V_{0,1}$ & $\frac{4\pi}{\beta^{2}}$ \\
\end{tabular}
\label{our}
\end{table}

\end{document}